\documentclass[a4paper]{jpconf}
\usepackage{graphicx}

\begin{document}
\title{Non-monotonic behavior of multiplicity fluctuations}

\author{M Rybczy\'nski and Z W\l odarczyk}

\address{Institute of Physics, \'Swi\c etokrzyska Academy,
ul. \'Swi\c etokrzyska 15, PL - 25-406 Kielce, Poland}

\ead{wlod@pu.kielce.pl}


\begin{abstract}
We discuss recently measured event-by-event multiplicity
fluctuations in relativistic heavy-ion collisions. It is shown
that the non-monotonic behavior of the multiplicity fluctuations
as a function of collision centrality can be fully explained by
the correlations between produced particles.
\end{abstract}.


\section{Introduction}
Event-by-event fluctuations in heavy-ion collisions have been
recently measured both at CERN SPS and BNL RHIC (see a brief
review \cite{Mitch}). The data (usually used $F$ or $\Phi$
measures to quantify the fluctuations of transverse momentum)
which show a non-trivial behavior as a function of collision
centrality, have been theoretically discussed from very different
points of view, including complete or partial equilibration,
critical phenomena, string or cluster percolation, production of
jets. In spite of these efforts a mechanism responsible for the
fluctuations is far from being uniquely identified.

\indent In \cite{Rybczynski1} we have established connection of
the $\Phi$-measure with the fluctuations of multiplicity and
advocated that fluctuations of transverse momentum are practically
irrelevant for the considered problem, i.e., for the apparent
structure seen in collision centrality dependence. We argue
\cite{Mrow} that the non-trivial behavior of the so called
transverse momentum fluctuations (quantified by $F$ or $\Phi$ -
measure) can be fully explained by the multiplicity fluctuations.

\indent Recently, the NA49 Collaboration published very first data
on multiplicity fluctuations as a function of collisions
centrality \cite{Gazdzicki:2004ef}, see also \cite{Rybczynski} in
this volume. Unexpectedly, the ratio $Var(N)/\langle N \rangle$,
where $Var(N)$ is the variance and $\langle N \rangle$ is the
average multiplicity of negative particles, changes
non-monotonically when number of wounded nucleons grows. It is
close to unity at fully peripheral ($N_w \le 10$) and completely
central ($N_w \ge 250$) collisions but it manifests a prominent
peak at $N_w \approx 70$. The measurement has been performed at
the collision energy $158 \, AGeV$ in the transverse momentum and
pion rapidity intervals $(0.005,1.5)$ GeV and $(4.0,5.5)$,
respectively. The azimuthal acceptance has been also limited, and
about $20\%$ of all produced negative particles have been used in
the analysis.

\indent The aim of this paper is to show that the observed effect
(non-monotonic dependence of the ratio $Var(N)/<N>$ on the number
of projectile participants $N_{P}$, i.e., on the production volume
$V$) stem from the correlations between produced particles and the
correlations are of nuclear and electromagnetic interactions
origin.


\section{Correlations and fluctuations}

\indent Let us introduce the single-particle configuration
distribution function \cite{Balescu}

\begin{equation}
n_{1}({\bf r}_{1})=n
\end{equation}

\noindent where $n$ is (constant) density of particles.
Two-particle configuration distribution function:

\begin{equation}
n'_{2}({\bf r}_{1},{\bf r}_{2})=n^{2}\cdot n_{2}(|{\bf r}_{1}-{\bf
r}_{2}|)\equiv n^{2}\cdot n_{2}(r)
\end{equation}

\noindent is expressed by two-particle correlation function
$\nu_{2}(r)$ following:

\begin{equation}
n_{2}(r)=1+\nu_{2}(r)
\label{nu2r12}
\end{equation}

\noindent where, asymptotically for $r \rightarrow \infty$,
$\nu_{2}(r)=0$ and $n_{2}(r)=1$.

\noindent For the volume $V$ one gets

\begin{equation}
<N_{V}>=\int \limits_{V}d{\bf r}_{1}n_{1}({\bf r}_{1})=\int
\limits_{V}d{\bf r}_{1}n
\end{equation}

\begin{equation}
<N_{V}^{2}>=\int\limits_{V}d{\bf r}_{1}n_{1}({\bf
r}_{1})+\int\limits_{V}d{\bf r}_{1}\int\limits_{V}d{\bf
r}_{2}n'_{2}({\bf r}_{1},{\bf r}_{2})
\end{equation}

\noindent The variance equals:

\begin{eqnarray}
<N_{V}^{2}>-<N_{V}>^{2} & = & \int\limits_{V}d{\bf
r}_{1}n_{1}({\bf r}_{1}) {}
                                  \nonumber\\
                      & & {} +\int\limits_{V}d{\bf r}_{1}\int\limits_{V}d{\bf r}_{2}[n'_{2}({\bf r}_{1},{\bf r}_{2})-n_{1}({\bf r}_{1})n_{1}({\bf r}_{2})]
\end{eqnarray}

\noindent and the normalized variance is given as:

\begin{equation}
\frac{<N_{V}^{2}>-<N_{V}>^{2}}{<N_{V}>}=1+\int\limits_{V}d{\bf
r}\left[n \cdot n_{2}(r)-n\right]
\label{varn_int0}
\end{equation}

\noindent Substituting now Eq. (\ref{nu2r12}) to Eq.
(\ref{varn_int0}) one imediately leads to the formula:

\begin{equation}
\frac{<N^{2}>-<N>^{2}}{<N>}=1+n \int\limits_{V}d{\bf r} \nu_{2}(r)
\label{varn_int}
\end{equation}

\noindent which shows clearly that the normalized variance is
given by two-particle correlation function.

\begin{figure}[h]
\begin{minipage}{18pc}
\includegraphics[width=18pc]{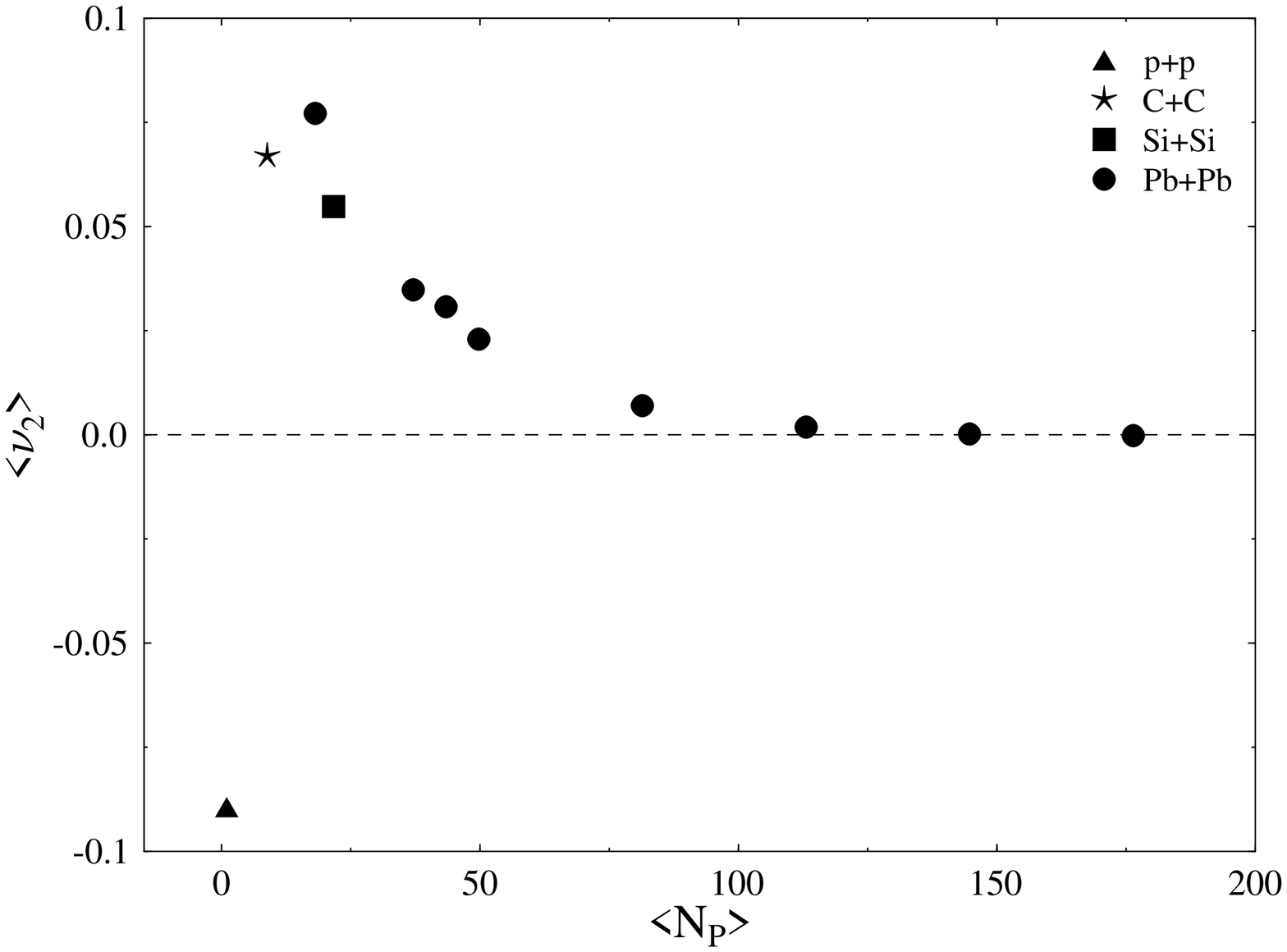}
\caption{\label{ni2_00N}Mean value of correlation function
calculated from data obtained at $158 \, AGeV$.}
\end{minipage}\hspace{2pc}%
\begin{minipage}{18pc}
\includegraphics[width=18pc]{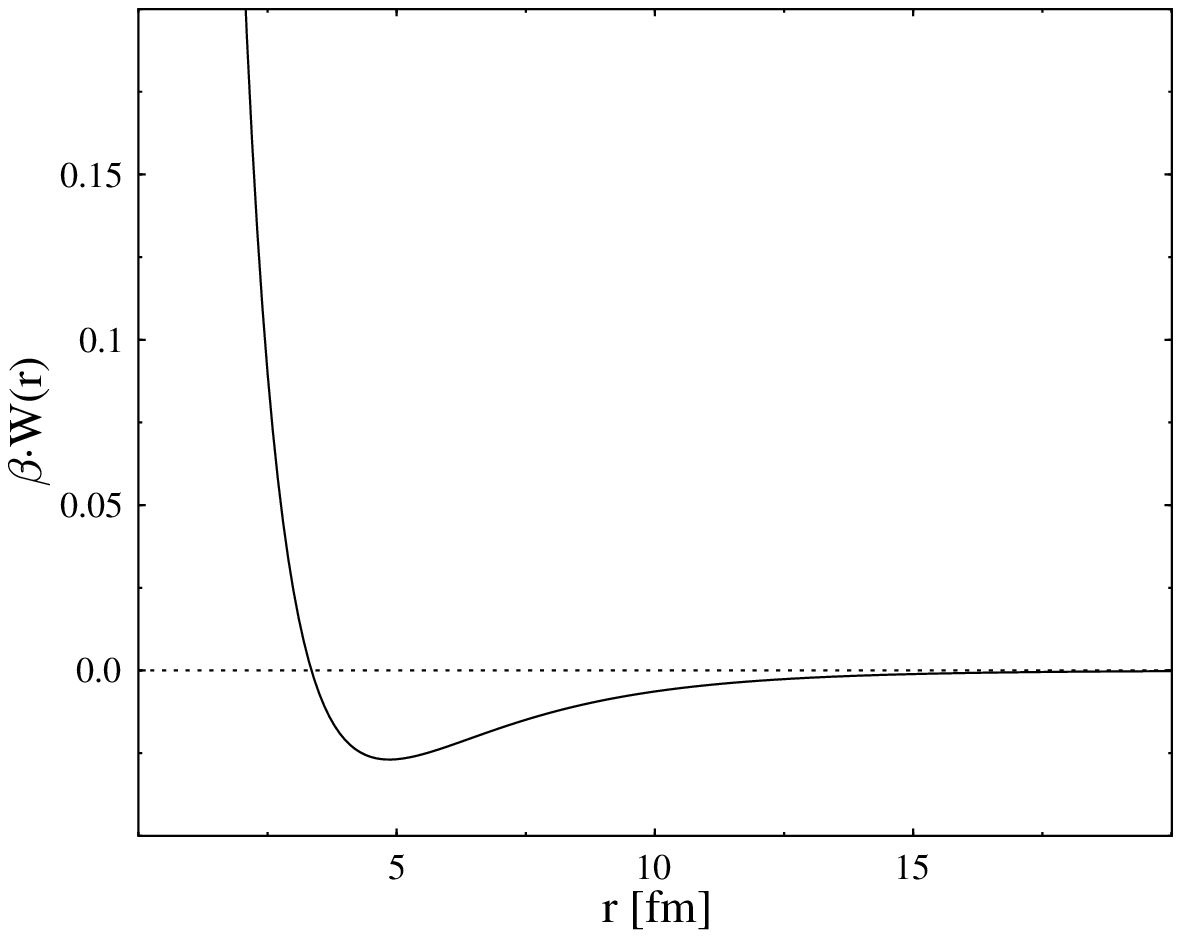}
\caption{\label{potencjal}The dependence of $\beta \cdot
W\left(r\right)$ on relative distance $r$.}
\end{minipage}
\end{figure}

\noindent Because:

\begin{equation}
\frac{Var(N)}{<N>}=1+nV<\nu_{2}>
\label{varn_nu2}
\end{equation}

\noindent where we note that the integral $\int\limits_{V}d{\bf
r}_{1}\int\limits_{V}d{\bf r}_{2} \nu_{2}(|{\bf r}_{1}-{\bf
r}_{2}|)=<\nu_{2}>V^{2}$, then it is clear that:

\begin{equation}
<\nu_{2}>=\frac{\left(\frac{Var(N)}{<N>}-1\right)}{<N>}
\label{nu2}
\end{equation}

\noindent has numerical values close to zero. Determined this way
$<\nu_{2}>$ do not depends on the detector acceptance $p$.
Because, the normalized variance for the accepted particles:

\begin{equation}
\frac{Var(N)}{<N>}=(1-p)+p\cdot \frac{Var(N_{p=1})}{<N_{p=1}>}
\end{equation}

\noindent and the accepted multiplicity:

\begin{equation}
<N>=p\cdot <N_{p=1}>
\end{equation}

\noindent one finds from Eq. (\ref{nu2}) the correlation function
for the accepted particles $<\nu_{2}>=<\nu_{2}>_{p=1}$. Fig.
\ref{ni2_00N} shows mean value of correlation function $<\nu_{2}>$
calculated from data on the ratio $Var\left(N\right)/<N>$ obtained
at $158 \, AGeV$ \cite{Gazdzicki:2004ef}.

\subsection{Simple example}

\indent Let the correlation function be in the form:

\begin{equation}
\nu_{2}({\bf r}_{1}-{\bf r}_{2})=\nu_{2}(r)=\exp(-2r/\lambda)
\end{equation}

\noindent In this case:

\begin{eqnarray}
\frac{Var\left(N\right)}{<N>}& = & 1+<N>\int\limits_{V}
\int\limits_{V}
d{\bf r}_{1}d{\bf r}_{2}\nu_{2}({\bf r}_{1}-{\bf r}_{2}){} \nonumber\\
                  & = &1+<N>\frac{2}{R^{2}}\int\limits_{0}^{R}(R-r)\nu_{2}(r)dr
\end{eqnarray}

\noindent and one gets

\begin{equation}
\frac{Var(N)}{<N>}=1+\frac{<N>}{2}\left(\frac{\lambda}{R}\right)^{2}\left(\exp\left(-2\frac{R}{\lambda}\right)
-1+2\frac{R}{\lambda}\right)
\end{equation}

\noindent It the limit $R>>\lambda$ we have
\begin{equation}
\frac{Var(N)}{<N>}\cong 1+<N>\frac{\lambda}{R}\cong 1
\end{equation}

\noindent then for $R<<\lambda$

\begin{equation}
\frac{Var(N)}{<N>}\cong 1+<N>
\end{equation}


\section{Correlations and interactions}

\indent Two-particle distribution function $n'_{2}(r)$ normalized
for mean density, in the equilibrium case (with inverse
temperature $\beta$) is given by the following Boltzmann factor
\cite{Balescu}:

\begin{equation}
n^{-1}n'_{2}(r)=n \cdot \exp\left(-\beta W(r)\right)
\end{equation}

\noindent For the rigid spheres with the radius $d_{0}$, where

\begin{equation}
W(r)=
\begin{array}{ll}
\infty & r<2d_{0}\\
0 & r>2d_{0}
\end{array}
\end{equation}

\noindent we have the correlation function:

\begin{equation}
\begin{array}{ll}
n_{2}(r)=0 & r<2d_{0}\\
\nu_{2}(r)=-1 & r<2d_{0}
\end{array}
\end{equation}


\subsection{Two potentials}

\indent Let us consider two potentials: the electrostatic Debye
potential \cite{Debye}

\begin{equation}
\varphi_{e}(r)=e\frac{\exp\left(-r/\lambda_{e}\right)}{r}
\end{equation}

\noindent and the nuclear Yukawa potential \cite{Yukawa}:

\begin{equation}
\varphi_{n}(r)=-g\frac{\exp\left(-r/\lambda_{n}\right)}{r}
\end{equation}

\noindent The attractive (nuclear) and repulsive (electrostatic)
interactions leads to the effective energy:

\begin{equation}
W\left(r\right)=e \cdot \varphi_{e}\left(r\right) - g \cdot
\varphi_{n}\left(r\right)
\end{equation}

\noindent The dependence of $\beta \cdot W\left(r\right)$ on
relative distance $r$ is presented in Fig. \ref{potencjal} for the
numerical values: $\lambda_{n}=1.5\,fm]$ and
$\lambda_{e}=3.5\,fm$.

\indent For the radius $r>2d_{0}$ one can write the correlation
function

\begin{equation}
\nu_{2}(r)=n_{2}(r)-1=\exp\left(a_{n}\frac{\exp\left(-r/\lambda_{n}\right)}{r}-a_{e}
\frac{\exp\left(-r/\lambda_{e}\right)}{r}\right)-1
\end{equation}

\noindent where strength factors are equal to: $a_{n}=\beta
g^{2}$, $a_{e}=\beta e^{2}=\left( 4\pi
n\lambda_{e}^{2}\right)^{-1}$ and interaction lengths are:
$\lambda_{n}=h\left(2\pi mc\right)^{-1}$, $\lambda_{e}=\left(4\pi
e^{2}n\beta\right)^{-1/2}$.

\noindent In the first approximation one can write

\begin{equation}
\nu_{2}(r)\approx
a_{n}\frac{\exp\left(-r/\lambda_{n}\right)}{r}-a_{e}
\frac{\exp\left(-r/\lambda_{e}\right)}{r}
\label{corr_eq}
\end{equation}

\noindent what allows for analytical integration $n\int d{\bf r}
\nu_{2}(r)$ and analytical estimation of normalized variance
$Var(N)/<N>$.

\indent The problem of local and global fluctuations (the so
called finite-size effect) was originally discussed by Nicolas
\cite{Nicolas} for correlation function:

\begin{equation}
\nu_{2}(r)=a\frac{\exp\left(-r/\lambda\right)}{r}
\end{equation}

\noindent Integrating it (via the Fourier transformations method)
in the spherical volume $V$ with the radius $R$ gives
\cite{Gardiner}:

\begin{eqnarray}
f\left(R,\lambda,a\right)& = &\frac{Var(N)}{<N>}=1+n \int d{\bf r}
a \frac{\exp\left(-r/\lambda\right)}{r}
                                                                                     \nonumber\\
   & = & 1+6 \pi an \frac{\lambda^{5}}{R^{3}}\Biggl[1-\left(\frac{R}{\lambda}\right)^{2}+\frac{2}{3}\left(\frac{R}{\lambda}\right)^{3}
    -\exp\left(-2R/\lambda\right)\cdot\left(1+\frac{R}{\lambda}\right)^{2}\Biggr]
\end{eqnarray}

\noindent Asymptotically for small $R$ $(V=4/3\pi R^{3}<<
\lambda^{3})$ one gets:

\begin{equation}
f\left(R,\lambda,a\right)=\frac{Var(N)}{<N>}\propto 1+ \frac{8 \pi
an R^{2}}{5}
\end{equation}

\noindent whereas for large $R$ $(V=4/3\pi R^{3}>> \lambda^{3})$
one gets:

\begin{equation}
f\left(R,\lambda,a\right)=\frac{Var(N)}{<N>}\propto 4 \pi an
\lambda^{2}\left( 1- \frac{3}{2} \frac{\lambda}{R}\right)
\end{equation}

\noindent For our correlation function $\nu_{2}\left(r\right)$,
given by Eq. (\ref{corr_eq}), one immediately gets:

\begin{equation}
\frac{Var\left(N\right)}{<N>}=f\left(R,\lambda_{n},a_{n}\right)-f\left(R,\lambda_{e},a_{e}\right)
\end{equation}

\noindent In this paper the integral Eq. (\ref{varn_int}) has been
calculated via Monte-Carlo method, assuming the homogenous
spherical production volume, $V\sim N_{P}$, to be proportional to
the number of participants $N_{P}$. Fig.\ref{r_deb_yuk} shows
normalized variance of the multiplicity distribution and mean
multiplicity of negatively charged particles produced in
collisions at $158 \, AGeV$ fitted by the two potentials model. It
is interesting to note that different shapes of the production
region (with particle density distribution $n\left(r\right)$)
influence the presented results for the large number of
participants $N_{P}$ and allow us to obtain much more satisfactory
fit to experimental data.


\subsection{Single potential (dipole-dipole interactions)}

\indent Electrostatic correlations play an important role in
analysis of thermodynamic system \cite{Brown,Levin}. In following,
we analyze one more (speculative, but possible) scenario, which
result in non-monotonic behavior of fluctuations.

\begin{figure}[h]
\begin{minipage}{18pc}
\includegraphics[width=18pc]{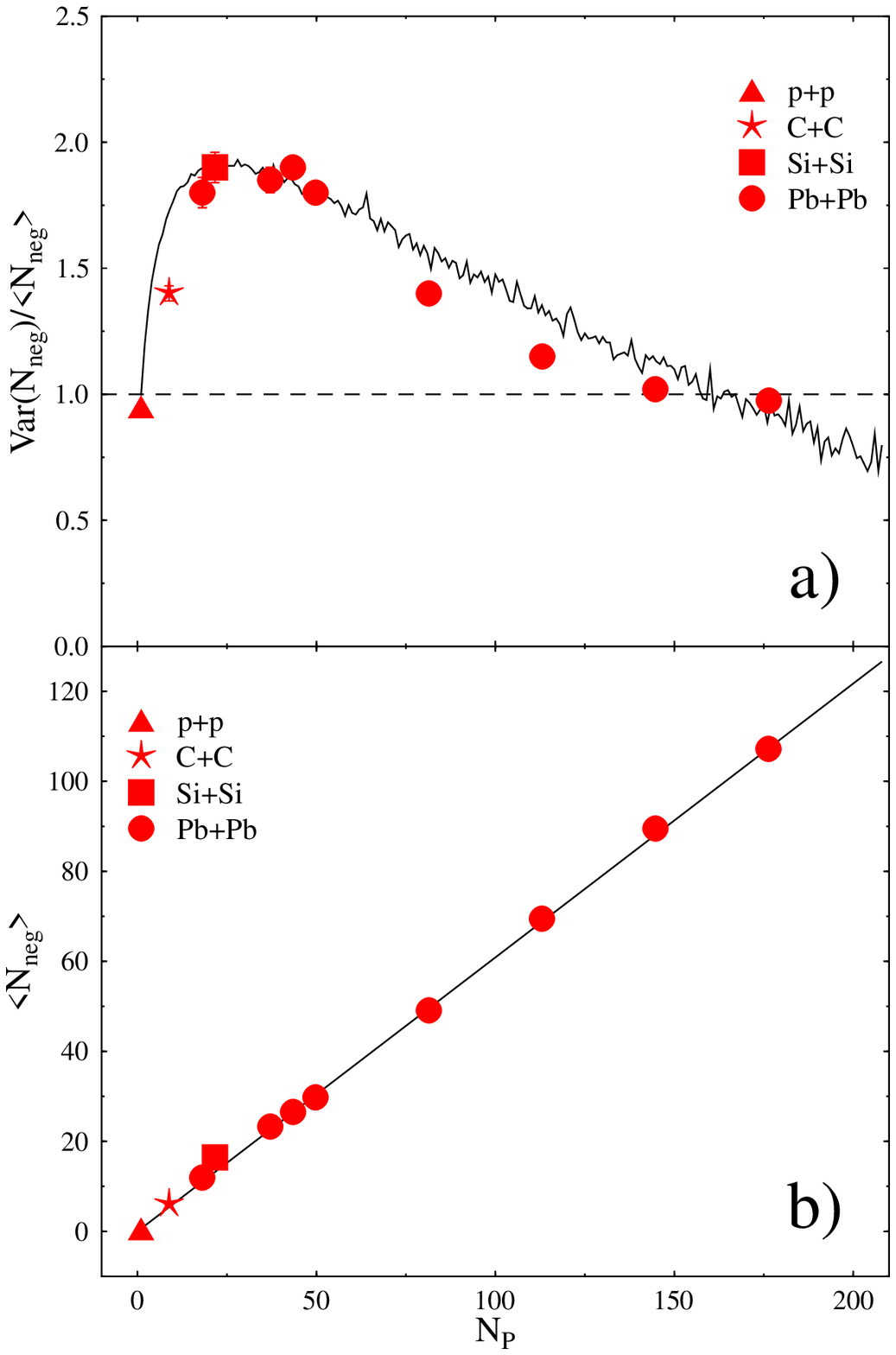}
\caption{\label{r_deb_yuk}Normalized variance of the multiplicity
distribution (a) and mean multiplicity (b) of negatively charged
particles produced in collisions at $158 \, AGeV$
\cite{Gazdzicki:2004ef} fitted by the two potentials model.}
\end{minipage}\hspace{2pc}%
\begin{minipage}{18pc}
\includegraphics[width=18pc]{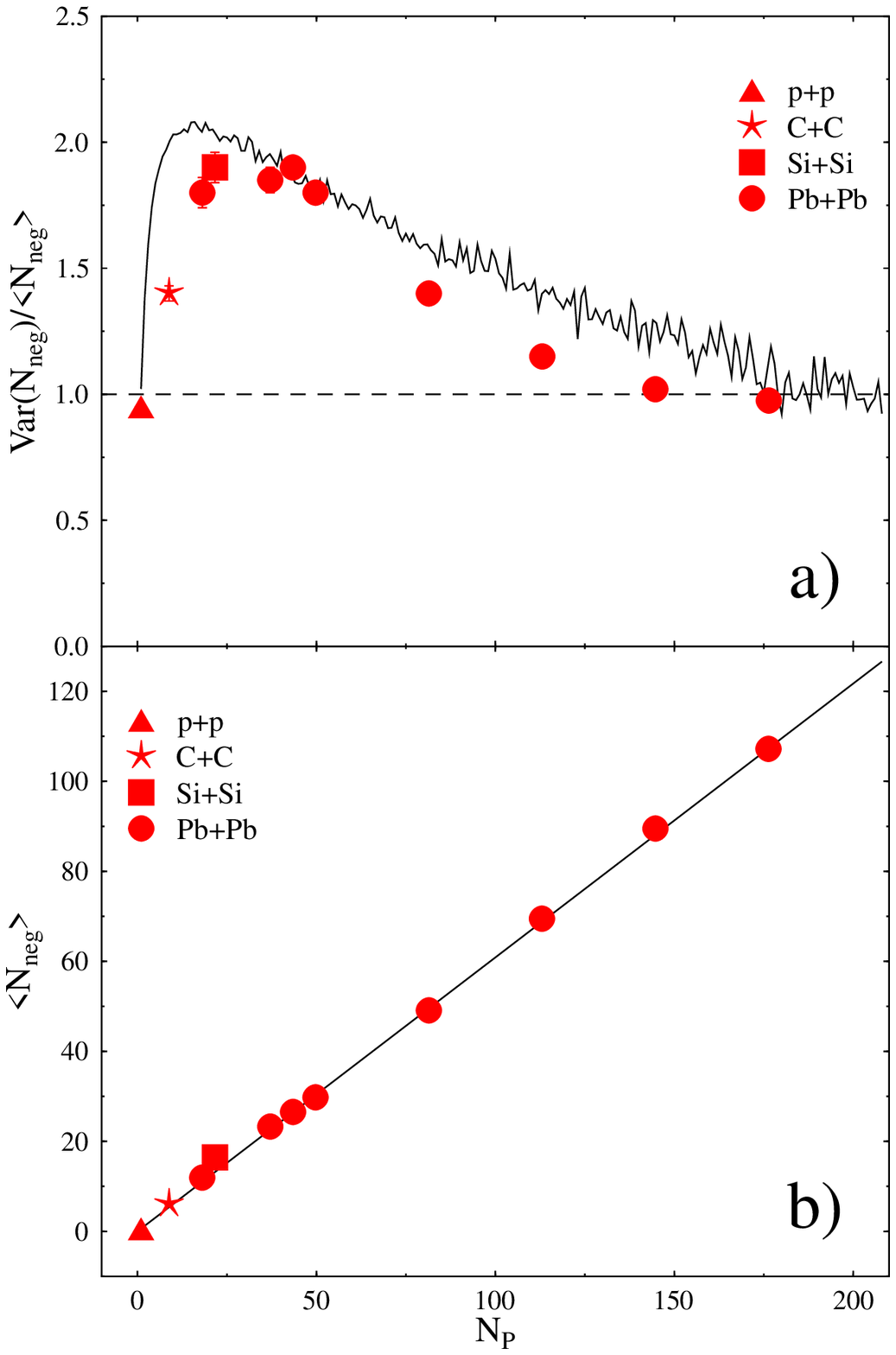}
\caption{\label{r_dipol}Normalized variance of the multiplicity
distribution (a) and mean multiplicity (b) of negatively charged
particles produced in collisions at $158 \, AGeV$
\cite{Gazdzicki:2004ef} fitted by the dipole model.}
\end{minipage}
\end{figure}

\noindent Let us assume that the particles are produced in pairs
$\left(\pi^{+}\pi^{-}\right)$ creating dipoles. The potential
describing interaction between ${\bf p}_{1}$ and ${\bf p}_{2}$
dipoles is given by:

\begin{eqnarray}
\varphi(r)& = &\frac{1}{r^{3}}\Biggl({\bf p}_{1}{\bf
p}_{2}-\frac{3\left({\bf p}_{1}{\bf r}\right)\left({\bf p}_{2}{\bf
r}\right)}{r^{2}}\Biggr)
                                       \nonumber\\
          & = &\frac{<p>^{2}}{r^{3}}\left(\sin\theta_{1}\sin\theta_{2}\cos\left(\phi_{1}-\phi_{2}\right)+\cos\theta_{1}\cos\theta_{2}-3\cos\theta_{1}\cos\theta_{2}\right)
                                        \nonumber\\
          & = &\frac{<p>^{2}}{r^{3}}f\left(\theta,\phi\right)
\end{eqnarray}

\noindent and changes from $-2<p>^{2}/r^{3}$ to $+2<p>^{2}/r^{3}$,
depending on dipoles space arrangement. Changing the polarization
of dipoles depending on distance between them
$f\left(\theta,\phi\right)\rightarrow f\left(r,\theta,\phi\right)$
allows for transition from negative values of potential (low
values of $r$; particles correlation) to positive (big $r$;
particles anti-correlation). For instance the dependence of the
type:

\begin{equation}
f\left(r,\theta,\phi\right)=-2\exp\biggl[-\left(r/\lambda\right)^{3}\biggr]+2\biggl(1-\exp\biggl[-\left(r/\lambda\right)^{3}\biggr]\biggr)
\end{equation}

\noindent provides to the correlation function $\nu_{2}(r)$ which
allow for approximate description of experimental data (c.f.
Fig.\ref{r_dipol}, where we fit data with the parameter
$\lambda=2.6\,fm$). One can get the effect of changing
$f\left(r,\theta,\phi\right)$ via correlation between the dipole
angles, for example:

\begin{equation}
\Delta\theta=\pi\biggl(1-\exp\biggl[-\left(r/\lambda\right)^{3}\biggr]\biggr)
\end{equation}

\section{Connections to percolation}

\indent The particles situated in correlation length $\lambda$
distance each other builds the cluster. For low $R<\lambda/2$ the
whole production volume belongs to the single cluster. If the size
of the fireball increases "the percolation" to the next clusters
appear \cite{Stauffer}. The number of percolation clusters $M_{C}$
is proportional to:

\begin{equation}
M_{C}\sim\left(\frac{2R}{\lambda}\right)^{3}
\end{equation}

\noindent The number of particle pairs in distance $d$, included
in the sphere with radius $R$ is distributed as:

\begin{equation}
P(d)=
\begin{array}{ll}
\frac{d}{R} & 0<d<R\\
2-\frac{d}{R} & R<d<2R
\end{array}
\end{equation}

\noindent and the number of particles pairs with relative distance
$d<\lambda$ to the total number of pairs is given by:

\begin{equation}
M_{P}=\frac{M_{P}\left(<\lambda\right)}{M_{TOT}}=
\begin{array}{ll}
2\left(\frac{\lambda}{2R}\right)^{2} & R>\lambda\\
4\left(\frac{\lambda}{2R}\right)-2\left(\frac{\lambda}{2R}\right)^{2}-1
& \frac{\lambda}{2}<R<\lambda\\
1 & R<\frac{\lambda}{2}
\end{array}
\end{equation}

\noindent In the simplest approximation the number of particle
pairs for single cluster equals $M_{P}/M_{C}$ and correlation
function $\nu_{2}\sim\left(M_{P}/M_{C}\right)^{2}$ is given by:

\begin{equation}
\nu_{2}=
\begin{array}{ll}
\Bigl(2\left(\frac{\lambda}{2R}\right)^{5}\Bigr)^{2} & R>\lambda\\
\Biggl(\left(\frac{\lambda}{2R}\right)^{3}\biggl[4\left(\frac{\lambda}{2R}\right)-2\left(\frac{\lambda}{2R}\right)^{2}-1\biggr]\Biggr)^{2}
& \frac{\lambda}{2}<R<\lambda\\
1 & R<\frac{\lambda}{2}
\end{array}
\end{equation}


\section{Connections to nonextensivity}

\indent Let us now continue discussion of the multiplicity
fluctuations from the point of view of its possible connection
with nonextensivity (for other hints on nonextensivity in hadronic
production processes and references to nonextensive statistics,
see \cite{Wilk, Navarra}). When there are only statistical
fluctuations in the hadronizing system one should expect the
Poissonian form of the multiplicity distribution. The existence of
intrinsic (dynamical) fluctuations result in broader distributions
usually expressed via Negative Binomial form and characterized by
the parameter (so called the inverse number of "clans"):

\begin{equation}
k^{-1}=\frac{Var(N)}{<N>^{2}}-\frac{1}{<N>}.
\end{equation}

\noindent By using Eq. (\ref{varn_nu2}),
$Var\left(N\right)/<N>=1+<N><\nu_{2}>$, then one gets:

\begin{equation}
q-1=k^{-1}=<\nu_{2}>
\end{equation}

\noindent where $q$ is an nonextensivity parameter.

\noindent Small systems (with diameters non exceeding correlation
length) are characterized by the parameter $q>1$ and just in the
thermodynamic limit $q=1$ (for big size systems with $R>>\lambda$,
where the correlation effects are small and all subsystems are
characterized by the same temperature).

\noindent The $q$ parameter could be connected with the number of
correlation clusters ("copies" of the system) $M_{C}$. If $M_{C}$
grows, the Tsallis entropy \cite{Tsallis}:

\begin{equation}
S=\lim_{q\to 1}\frac{\sum p^{q}-1}{1-q}
\end{equation}

\noindent where nonextensivity parameter $q=1+m\sim
1+\frac{1}{M_{C}}$ leads to the thermodynamic Shannon entropy:

\begin{equation}
S=-\sum p \ln p
\end{equation}

\noindent because:

\begin{equation}
\ln p=\lim_{m\to 0}\frac{p^{m}-1}{m}.
\end{equation}

\noindent Experimental data prefers dependence of the type:

\begin{equation}
M_{C}\sim \exp\Biggl(\left(\frac{R}{\lambda}\right)^{3}\Biggr).
\end{equation}


\section{Conclusions}
The observed effect (non-monotonic dependence of the ratio
$Var(N)/<N>$ on production volume $V\sim N_{P}$) stem from the
correlations between produced particles. The correlations are of
nuclear and electromagnetic interactions origin. It is possible to
describe the appearance of a prominent peak in $Var(N)/<N>$ at
number of participants $N_{P} \approx 35$, for charged particles.
For the neutral particles we observe lack of non-monotonic
behavior of $Var(N)/<N>$ when number of wounded nucleons grows
(satisfy the experimental observation \cite{WA98}). Effect depends
on energy, because particles density $n$ changes with energy (the
magnitude of the effect $\simeq n$ and position of maximum also
shifts, because correlation length $\lambda_{e}\simeq
1/\sqrt{n}$).


\end{document}